\documentclass[preprint,amsmath,amssymb,aps,showkeys]{revtex4-2}
\usepackage{graphicx}
\usepackage{dcolumn}
\usepackage{bm}
\usepackage{amsmath}
\usepackage{amssymb}
\usepackage{textcomp}
\usepackage{graphicx}
\usepackage{caption}
\usepackage{subcaption}
\usepackage{array}
\usepackage{mathtools}
\usepackage{centernot}
\usepackage{slashed}
\usepackage{hyperref}
\usepackage{xcolor}

\begin{document}
\title{Triplet Higgs assisted leptogenesis from axion oscillation after inflation}
\author{Sasmita Mishra}
\email{mishras@nitrkl.ac.in}
\affiliation{%
Department of Physics and Astronomy, National Institute of
Technology Rourkela, Sundargarh, Odisha, India, 769 008}%
\date{\today}
	
\begin{abstract}
Leptogenesis via axion oscillation after inflation is an alternate
mechanism of thermal leptogenesis. In this mechanism, the requirement of 
the existence of a lepton number ($L$) violating process in equilibrium 
to drive the lepton number requires the temperature of leptogenesis
at $\sim 10^{13}$ GeV. Triplet scalars, due to their interaction with gauge
bosons, make a suitable candidate to prevail in the thermal bath via 
gauge scattering at such high energy. Also, owing to its interaction with Standard Model (SM)
leptons and Higgs scalar, it can mediate $\Delta L =2$ process. Moreover,
just one triplet is enough to serve the purpose as opposed to thermal 
leptogenesis where at least one more triplet scalar/right-handed neutrino
is required to generate a sizable $CP$ violation. In this work, we study 
a model where the SM gauge group is extended with $U(1)_{\rm PQ}$, with 
the addition of one Higgs doublet, one scalar triplet, and one complex scalar singlet. The presence of a complex scalar singlet decouples the PQ symmetry breaking from the 
electroweak scale. It also provides a common source of 
an axion-like particle and seesaw scale. The scalar 
triplet offers a common link between leptogenesis via axion oscillation and neutrino mass. Further with the presence of one triplet scalar, the lepton flavor violation process is directly determined from low-energy neutrino oscillation data.

\end{abstract}
	
	
\maketitle
\newpage   
\section{Introduction}
\label{sec:intro}
The origin of the baryon asymmetry of the Universe (BAU) after 
 an early epoch of cosmological inflation continues
to be one of the most important questions of physics beyond the SM. The observations from cosmic microwave background (CMB) radiation and big bang nucleosynthesis give a measurement of the asymmetry \cite{Planck:2018vyg}, $Y_B=\frac{(n_B-n_{\overline{B}})}{s}$ as  $8.7\times 10^{-11 }$, where $n_{B}$, $n_{\overline{B}}$, and $s=\frac{2\pi^2}{45}g_{*s}T^{3}$ denote the number densities of baryons, antibaryons and entropy density of the Universe respectively. Baryogenesis via leptogenesis \cite{Fukugita:1986hr} has become a well-known scenario to explain the matter-antimatter asymmetry of the Universe based on the Sakharov's conditions \cite{Sakharov:1967dj}. The conditions provide the necessary
ingredients for dynamical generation of BAU viz; baryon(lepton) number
($B(L)$)
 violation, $C$ and $CP$ violation and out-of-equilibrium dynamics in the processes involving baryons. The well-known seesaw framework could explain tiny non-zero neutrino mass and the consequent leptogenesis leading to adequate baryon asymmetry. 
One such scenario is the SM augmented with one scalar triplet where through 
type-II seesaw mechanism  \cite{Magg:1980ut, Lazarides:1980nt, Mohapatra:1980yp, Schechter:1980gr} 
light neutrino masses can naturally arise from Weinberg's dimension-5 operator.
The unique feature of type-II seesaw-induced light neutrino mass
is that  the presence of only one 
Yukawa coupling matrix can completely be
determined from the low-energy values of neutrino masses
and mixing angles up to an overall factor and renormalization group
equation (RGE) evolution effects \cite{Rossi:2002zb}.
With one triplet added to the SM, although it is 
possible to theoretically explain the light neutrino masses and mixing
satisfying experimental data, thermal leptogenesis
is not viable.
It was shown in Ref. \cite{Fry:1980bc} baryogenesis via heavy Higgs leptogenesis requires its mass $\ge 10^{13}$ GeV with the amount of CP violation $\sim 10^{-7}$. However to generate CP violation $\sim 10^{-7}$
one triplet scalar is inadequate, which leads to the inclusion of another right-handed neutrino or triplet scalar. 
As a result the opportunities to
connect high- and low-scale dynamics is lost in a model.
To preserve the intriguing connection between high- and low-scale dynamics, in their work \cite{Barrie:2021mwi,Barrie:2022cub}, the authors study
with a single triplet scalar added to the SM, it can
simultaneously give rise to BAU and a Starobinsky-like inflationary
epoch \cite{Starobinsky:1980te}. In this work, we show an alternate mechanism: where without
losing the direct link between neutrino mass and flavor structure
leptogenesis can be achieved with the help of one triplet scalar using 
axion oscillation after inflation \cite{Kusenko:2014uta}. 

The quantum-chromodynamics (QCD) axion and/or axion-like-particles (ALPs) 
appear from a spontaneously broken global $U(1)_{\rm PQ}$ symmetry 
as a pseudo-Nambu-Goldstone boson, the phase of a complex singlet scalar field. 
The simplest realizations of PQ mechanism in Weinberg-
Wilczek (WW) model \cite{Weinberg:1977ma, Wilczek:1977pj}.
The WW model was soon ruled out by laboratory searches as the 
axion decay constant $f_a$ is of the order of the electroweak scale,
$v_{\rm ew} \simeq 246$ GeV, not being sufficiently suppressed. 
This led to models, in which the PQ symmetry breaking is decoupled from the electroweak scale via the introduction of an SM singlet scalar field,
$\sigma$ acquiring a vacuum expectation value (VEV) $f_a$. 
In all string theories, axions and ALPs appear naturally as Kaluza-Klein zero modes of antisymmetric tensor fields when the extra six dimensions are compactified. Determined by the anomaly cancelation conditions the couplings of these fields result in the anomalous CP violating couplings in low-energy effective Lagrangian. Thus axions and ALPs are a potential bridge connecting the low-energy effective theories like SM and high-energy theories like string theory, theory of extra dimension, UV complete theories like grand unified theories (GUT), and more. ALPs have wide applications in theoretical physics. Some of them remain relatively
light and play an important role in baryogenesis, the role of inflaton, dark matter, and dark energy (for a review see Ref. \cite{DiLuzio:2020wdo}).
In this work, we explore one such model that contains ALP and additional Higgs scalars in the SM to address the issue of neutrino mass and baryogenesis. 
We intend to study the common origin of 
neutrino mass, and baryogenesis through spontaneously broken PQ symmetry in a way that  the PQ symmetry breaking scale and the seesaw
scale arises from a common origin.
A related study of the common origin of resolving the strong CP problem and the seesaw mechanism explaining the smallness
of neutrino masses with spontaneously broken PQ symmetry
was studied in Ref. \cite{Ahn:2015pia}.

Sakharov's conditions for attaining successful baryogenesis are based on
$CPT$ invariance.  In an attempt by Cohen and Kaplan in their scenario of spontaneous baryogenesis \cite{Cohen:1988kt}, the authors have shown that in case $CPT$ is spontaneously broken baryon asymmetry is generated in thermal equilibrium. In an attempt in a similar 
direction it was shown in Ref. \cite{Co:2019wyp} that the PQ symmetry may be explicitly broken in the early universe, inducing the rotation of
PQ charged scalar field. The asymmetry of the PQ charge arising from rotation can be converted to baryon asymmetry via QCD and electroweak sphaleron transitions.
In Ref. \cite{Duch:2025abl}, it was shown baryogenesis from cosmological evolution of a $CP$-
breaking modulus scalar either during the Big Bang or around the end of inflation with a condition that modulus scalar evolves in time while $\Delta L = 2$ rates, related to neutrino masses, go out of equilibrium at a decoupling temperature $T_{\rm dec} \sim 10^{11}$ GeV.

In this work, we pursue the idea of the possibility of 
baryogenesis via leptogenesis by spontaneously breaking  of 
$CPT$ invariance by introducing an ALP, which couples derivatively to the fermion current $j_\mu$ in the Lagrangian. The presence of the derivative 
coupling induces an effective chemical potential for the fermion number
shifting the energy levels of fermions and antifermion concerning
each other. This amounts to creating the difference between $B$ and $L$
satisfying one of Sakharov's conditions. After the end of inflation, 
the ALP starts to oscillate coherently with nonzero velocity which temporarily induces an
effective chemical potential for the fermion number. For baryogenesis, there must be an external
source of baryon or lepton number violation in equilibrium. 
The reheating temperature in this scenario turns out to be $\simeq10^{13}$
GeV makes it a high-energy scale phenomenon, making it an alternate mechanism of thermal leptogenesis. Inert Higgs doublet assisted lepton number violating 
operator with a much lower reheating temperature was studied in 
Ref. \cite{Datta:2024xhg} via axion oscillation. In this work we propose the presence
of a single triplet scalar is suitable enough to assist the 
$\Delta L =2$ process and generate light neutrino mass via type-II
seesaw mechanism, making it a unified picture.
While the involvement of gauge coupling in 
case of triplet scalars ensures its presence in the thermal bath at an early Universe, the PQ symmetry breaking is decoupled from the
electroweak scale via the introduction of an SM singlet scalar field,
$\sigma$ acquiring a vacuum expectation value (VEV) $f_a$.
 
The paper is organized as follows. In section (\ref{sec:model}), 
we lay down the model where the SM is extended with $U(1)_{\rm PQ}$
symmetry with the addition of one Higgs doublet, one complex scalar singlet and one triplet scalar.
In section(\ref{sec:lgnss}), we discuss the triplet scalar-assisted 
leptogenesis via axion oscillation. In section(\ref{sec:lfv}) we discuss
the lepton flavor violation by connecting the branching ratios
with light neutrino mass matrix through triplet Yukawa coupling. The conclusions are given in section (\ref{sec:concl}).
\section{The Model}
\label{sec:model}
Although the presence and abundance of axions are linked with 
an effective low-energy consequence of string theory, the theory also provides strong evidence that exact
global symmetries cannot survive in contact with quantum gravity. 
Also, obtaining an automatic $U(1)_{\rm PQ}$ in  GUT or connecting the axion and grand unification is a longstanding problem \cite{DiLuzio:2020qio}. This
is similar to the so-called PQ quality problem \cite{Georgi:1981pu}: why is $U(1)_{\rm PQ}$ a good symmetry of ultraviolet (UV) physics? Addressing this issue, 
simultaneous presence of horizontal flavor group $SU(3)_f$ and vertical $SO(10)$ leads to an automatic $U(1)_{\rm PQ}$ if the $SO(10)$ Higgs representations are properly chosen was shown in Ref. \cite{Georgi:1981pu} with a global $SU(3)_f$ and local
$SU(3)_f$ in Ref. \cite{DiLuzio:2020qio}. Some related proposals without GUTs were carried out in references \cite{Chang:1984ip},  \cite{Jaeckel:2010ni} and \cite{Pal:1994ba}.

In this section, we present a model motivated by the idea of the connection of massive neutrinos, often related to a high-energy scale 
seesaw mechanism, 
with the presence of spontaneously broken PQ symmetry \cite{Bertolini:2014aia}. 
For UV completion of the theory we choose 
Peccei and Quinn symmetry through $U(1)_{\rm PQ}$ symmetry.
The gauge group of the model is the SM gauge
group extended by $U(1)_{\rm PQ}$ with the addition of a complex scalar
field $\sigma$ and a Higgs doublet. The addition of $\sigma$ 
decouples the PQ symmetry breaking from the electroweak scale.
The particle content of the model is given in table (\ref{tab:particle}) with the corresponding charges. The 
determination of PQ charges is summarised in the appendix
(\ref{app:PQcharge}).
\begin{center}
\begin{tabular}{|c| c |c| c |c |c|}
\hline
Field & Spin & $SU(3)_c$ & $SU(2)_L$ & $U(1)_Y$ & $U(1)_{\rm PQ}$ \\ \hline \hline
$q_L$ & $\frac{1}{2}$ & $3$ & $2$ & $+\frac{1}{6}$ & 0\\
$u_R$ & $\frac{1}{2}$ & $3$ & $1$ & $+\frac{2}{3}$ & $X_u$\\
$d_R$ & $\frac{1}{2}$ & $3$ & $1$ & $-\frac{1}{3}$ & $X_d$\\ \hline 
$l_L$ & $\frac{1}{2}$ & $1$ & $2$ & $-\frac{1}{2}$ & $X_l$\\
$e_R$ & $\frac{1}{2}$ & $1$ & $1$ & $-1$ & $X_e$\\ \hline 
$H_u$ & $0$ & $1$ & $2$ & $-\frac{1}{2}$ & $-X_u$\\
$H_d$ & $0$ & $1$ & $2$ & $+\frac{1}{2}$ & $-X_d$\\ \hline 
$\Delta$ & $0$ & $1$ & $3$ & $+1$ & $X_\Delta$\\ \hline
$\sigma$ & $0$ & $1$ & $1$ & $0$ & $X_\sigma$\\
\hline
\end{tabular}
\label{tab:particle}
\end{center}
The Lagrangian for the fermion sector is given by,
\begin{equation}
 -\mathcal{L} = Y_u \bar{q}_L u_R H_u + Y_d \bar{q}_L d_R H_d
 + Y_e \bar{l}_L e_R H_d + 
 \frac{1}{2}Y_\Delta l_L^T C i \tau_2 \Delta l_L
 + {\rm h.c.},
 \label{eq:lagrng}
\end{equation}
where the triplet scalar, $\Delta$ can be represented as,
\begin{equation}
 \Delta \equiv \frac{ \vec{\tau}\cdotp \vec{\Delta}}{\sqrt{2}}=
 \begin{pmatrix}
  \frac{\Delta^+}{\sqrt{2}} & \Delta^{++}\\
  \Delta^0 & -\frac{\Delta^+}{\sqrt{2}}
 \end{pmatrix},
\end{equation}
where $\vec{\Delta} = (\Delta_1, \Delta_2, \Delta_3)$ are $SU(2)_L$
components of the scalar triplet.
The scalar potential is given by,
\begin{equation}
 V = V(\sigma) + V(\Delta ) + V(H_u, H_d) + V (\sigma, H_u, H_d)
 +V(\sigma, \Delta)+ V(\Delta, H_u, H_d)+ V(\sigma, \Delta, H_u, H_d),
\end{equation}
where the individual terms are given as 
\begin{eqnarray}
\nonumber
 V(\sigma) &=& \lambda\left( \sigma^* \sigma -\frac
 {f_a^2}{2}\right)^2,\\
 \nonumber
  V(\Delta ) &=& \mu_\Delta^2 {\rm Tr} \left(\Delta^\dagger \Delta \right) +\lambda_{\Delta 4} 
{\rm Tr} \left(\Delta^\dagger \Delta \right){\rm Tr} \left(\Delta^\dagger \Delta \right)+\lambda_{9} {\rm Tr}
\left(\Delta^\dagger \Delta \right)^2, \\ \nonumber
 V(H_u, H_d) &=& -\mu_1^2 |H_u|^2 + \lambda_1|H_u|^4 - \mu_2^2 |H_d|^2
 +\lambda_2 |H_d|^4 +\lambda_{12} |H_u|^2 |H_d|^2 +\lambda_4 |H_u^\dagger H_d|^2,\\
 \nonumber 
 V (\sigma, H_u, H_d) &=& \lambda_{13} |\sigma|^2 |H_u|^2
 + \lambda_{23} |\sigma|^2 |H_d|^2 + \left( \lambda_5 \sigma^2 \tilde{H}_u^\dagger H_d  
+ {\rm h.c.}  \right), \\ \nonumber
 V(\sigma, \Delta) &=& \lambda_{\Delta 3} |\sigma|^2 {\rm Tr} \left(\Delta^\dagger \Delta \right),\\ \nonumber
 V(\Delta, H_u, H_d) &= &  {\rm Tr} \left(\Delta^\dagger \Delta \right)
\left[ \mu_{\Delta 1} |H_u|^2+ \mu_{\Delta 2} |H_d|^2 
 \right] + \lambda_{7}H_u^\dagger \Delta \Delta^\dagger H_u
+\lambda_{8}H_d^\dagger \Delta \Delta^\dagger H_d +\lambda_{9}
\left(\Delta^\dagger \Delta \right)^2, \\ \nonumber
 V(\sigma, \Delta, H_u, H_d) &=& \lambda_6 \sigma H_u^\dagger \Delta^\dagger H_d + {\rm h.c.},
\end{eqnarray}
where $\tilde{H}_u = i \tau_2 H_u^*$.
The simultaneous presence of $Y_\Delta$ and $\lambda_6$ ensures explicit breaking of the lepton number.

The vacuum configurations of the PQ-charged complex singlet
is given by,
\begin{equation}
 \sigma = \frac{1}{\sqrt{2}}(S+f_a) e^{i\phi_a/f_a}. 
\end{equation}
At the first stage of symmetry breaking with the 
radial component of the scalar singlet 
$\sigma$ develops a VEV, $f_a$ 
and breaks the PQ symmetry. The phase of the complex scalar field 
$\theta = \phi_a/ f_a$, where the pseudo-Numbu-Goldstone boson, 
$\phi_a$ is identified as the ALP.
The massive radial mode is settled to
the minimum and ignored
below the scale $f_a$. Assuming that the PQ symmetry is broken before the end of inflation, the initial value of the ALP $\phi_{a0} = \theta_0 f_a$,
where the value of the angle $\theta_0$ takes a random value in the range $\theta_0 \in [0, 2\pi)$. So the initial value of $\phi_{a0}$ results in
being constant during the superhorizon scale. The potential of the ALP 
can originate from 
instanton effects in a strongly coupled hidden sector characterized by
a dynamical scale $\Lambda_H$ and is given by,
\begin{equation}
V(\theta)
 = \Lambda_H^4 \left( 1- \cos\theta \right).
 \label{eq:axion-pot}
\end{equation}
The potential can be approximated to $V(\theta)\simeq \frac{1}{2} 
m_a^2 \phi_a^2$, where the ALP mass $m_a \simeq \Lambda_H^2/f_a$.
After the PQ symmetry breaking the potential can be written as,
\begin{eqnarray}
 V_1 &=& V(\theta)+V_{H_u,H_d} + 
 \lambda_{13} f_a^2 |H_u|^2
 + \lambda_{23} f_a^2 |H_d|^2  \\ \nonumber
&+ &  {\rm Tr} \left(\Delta^\dagger \Delta \right)
\left[\mu_\Delta^2 + \mu_{\Delta 1} |H_u|^2+ \mu_{\Delta 2} |H_d|^2 
 + \lambda_{\Delta 3} f_a^2 +  \lambda_{\Delta 4} 
{\rm Tr} \left(\Delta^\dagger \Delta \right)\right] \\ \nonumber
&+& \lambda_{7}H_u^\dagger \Delta \Delta^\dagger H_u
+\lambda_{8}H_d^\dagger \Delta \Delta^\dagger H_d +\lambda_{9}
\left(\Delta^\dagger \Delta \right)^2 \\ \nonumber
&+& \left( \lambda_5 f_a \tilde{H}_u^\dagger H_d  
+\lambda_6 f_a H_u^\dagger \Delta^\dagger H_d + {\rm h.c.}\right).
\end{eqnarray}
An the first stage of $U(1)_{\rm PQ}$ symmetry breaking, when the electroweak and triplet scalar VEVs are not switched on ($\langle H_{u,d} \rangle =0 $), the triplet scalar $\Delta$ acquires a bare
mass $\mu_\Delta$ as well as a mass induced by the VEV of the complex singlet scalar charged 
under $U(1)_{\rm PQ}$.
The model now can be treated as a two-Higgs-doublet model (2HDM) extended with a real triplet scalar. For a reference see Ref.        \cite{Ait-Ouazghour:2020slc}. In this case the scalar spectrum
can be calculated and is summarised in the appendix (\ref{sec:scalar}). By electroweak and triplet scalar VEVs switched on,
the scalar spectrum has been calculated in Ref. \cite{Bertolini:2014aia}).
\subsection{Neutrino mass}
After the first stage of symmetry breaking, at a later epoch the 
remaining scalar fields attain the vacuum configuration given by,
\begin{equation}
 H_u \supset 
 \frac{v_u+h_u}{\sqrt{2}} 
 \begin{pmatrix}
  1\\
  0
 \end{pmatrix},
 ~~ H_d \supset \frac{v_d+h_d}{\sqrt{2}} 
  \begin{pmatrix}
  0\\
 1
 \end{pmatrix},~~
  \Delta \supset \frac{v_\Delta + \delta^0} {\sqrt{2}} 
  \begin{pmatrix}
   0 & 0\\
   1 & 0 
  \end{pmatrix},
  \label{eq:vac-allign}
\end{equation}
where $v_u, v_d$ and $v_\Delta$ denote the 
corresponding VEVs.
The light neutrino mass matrix from Eq.(\ref{eq:lagrng})
can be written as,
\begin{equation}
 m_\nu = Y_\Delta v_\Delta.
 \label{eq:nu-mass}
\end{equation}
Using the VEVs of $H_u, H_d$ and $\Delta$ from Eq.(\ref{eq:vac-allign}), the part of the potential dependent on $v_\Delta$ is
given as,
\begin{eqnarray}
\nonumber
 \langle V_\Delta \rangle &=& \left( \mu_\Delta^2 +
 \lambda_{\Delta 3} f_a^2 + \lambda_{\Delta 1} v_u^2
 +(\lambda_{\Delta 2} +\lambda_{8})v_d^2\right)v_\Delta^2\\ \nonumber
 &+& 2 \lambda_6  f_a v_u v_d v_\Delta +{\mathcal O}(v_\Delta^4)
 + {\rm terms ~ independent ~ of} ~v_\Delta.
\end{eqnarray}
For a hierarchy $f_a \gg v_{u,d}\gg v_\Delta$, the minimization
of the scalar potential is given by,
\begin{equation}
 2 M_\Delta^2 v_\Delta + 2 \lambda_6 f_a v_u v_d \approx 0,
\end{equation}
\begin{equation}
 M_\Delta^2 = \mu_\Delta^2 + \lambda_{\Delta 3} f_a^2 + \lambda_{\Delta 1}
 v_u^2 +(\lambda_{\Delta 2} + \lambda_{\Delta 8})v_d^2.
\end{equation}
Writing $\mu = \lambda_6 f_a$ and $v_u v_d \approx v_{\rm ew}^2$, Eq.(\ref{eq:nu-mass}) 
can be written as
\begin{equation}
 m_\nu = Y_\Delta v_\Delta \approx \frac{Y_\Delta \mu v^2_{\rm ew}}
 {M_\Delta^2}; ~~~ v_\Delta \approx \frac{\lambda_6 f_a v_u v_d}
 {M_\Delta^2}.
 \label{eq:nu-mass-std}
\end{equation}
The VEV $v_\Delta$ has the following range of allowed values,
\begin{equation}
 {\mathcal{ O}}(1) {\rm GeV} > v_\Delta \gtrsim 0.05 {\rm eV },
 \end{equation}
where the upper bound is derived from the $\rho$-parameter constraints determined from precision measurements \cite{Kanemura:2012rs}, and the lower limit ensures the generation of
the observed neutrino masses while also requiring perturbative Yukawa couplings.
In the limit $\mu \approx M_\Delta$ and $v_{\rm ew} \sim {\mathcal{O}}(100)$ GeV, the mass of
the triplet scalar is constrained as
\begin{equation}
10^4~{\rm GeV} \le  M_\Delta \le 10^{13}~{\rm GeV}. 
\end{equation}
The above bound is obtained by considering that there should be at least one neutrino mass at least of the
order of the $m_\nu \sim 0.05$ eV. By diagonalizing the mass matrix $m_\nu = Y_\Delta v_\Delta$ with a 
diagonalizing matrix $U_L$
\begin{equation}
 U_L^T m_\nu U_L = m_{\rm Diag},  
\end{equation}
the eigen values of the light neutrino masses $m_i, i= 1,2,3$ can be obtained.
In the present work, we have taken $U_L$ as the
Pontecorvo-Maki-Nakagawa-Sakata (PMNS) matrix, $U_{\text{PMNS}}$ 
times a diagonal matrix involving Majorana phases, 
${\rm Diag}(1, e^{i\alpha}, e^{i\beta})$. The structure 
of $U_{\rm PMNS}$ matrix can be formed using three rotation 
matrices with corresponding rotating angles $\theta_{ij};~ i,j =1,2,3$,
and a Dirac CP violating phase $\delta_{CP}$. The PMNS matrix
for light neutrino sector is,
	\begin{equation}
	\centering
	U_{\rm PMNS} = 
	\begin{pmatrix} 
	c_{12}c_{13} & s_{12}c_{13} & s_{13} e^{-i\delta_{CP}}  \\
	-s_{12}c_{23} - c_{12}s_{13}s_{23}e^{i\delta_{CP}} & c_{12}c_{23} - s_{12}s_{13}s_{23}e^{i\delta_{CP}} & c_{13}s_{23} \\ s_{12}s_{23} - c_{12}s_{13}c_{23}e^{i\delta_{CP}} & -c_{12}s_{23} - s_{12}s_{13}c_{23}e^{i\delta_{CP}} & c_{13}c_{23}
	\end{pmatrix},                  
	\end{equation}
where $c_{ij} = \cos (\theta_{ij})$, $s_{ij} = \sin (\theta_{ij})$.
\section{Leptogenesis from axion oscillation }
\label{sec:lgnss}
The axion is the angular direction of the complex scalar field, $\sigma$.
At the first stage of symmetry breaking with the 
radial component of the scalar singlet 
$\sigma$ develops a VEV, $f_a$ 
and breaks the PQ symmetry. The massive radial mode is settled to
the minimum heavy and ignored
below the scale $f_a$. Assuming that the PQ symmetry is broken before the end of inflation, the initial value of the axion at the end of inflation $\phi_{a0} = \theta_0 f_a$,
where the value of he angle $\theta_0$ takes a random value in the range $\theta_0 \in [0, 2\pi)$. So the initial value of $\phi_{a0}$ results in
being constant during the superhorizon scale. The evolution of the axion field in the early universe is governed by the classical field equation
\begin{equation}
 \ddot{\phi_a} +3H \dot{\phi_a} = -\partial_{\phi_a} V\left( \theta \right),
\end{equation}
where $V\left( \theta \right)$ is given in Eq.(\ref{eq:axion-pot}). 
The axion mass $m_a \simeq \Lambda_H^2/f_a$ can originate from
instanton effects in a strongly coupled hidden sector characterized by
a dynamical scale $\Lambda_H$.
For compatibility, $m_a$ should be smaller than the scale of inflation $H_{\rm inf}$, $m_a\lesssim H_{\rm inf}$. 
After the end of inflation, the axion field remains at rest until the  
Hubble parameter drops to $H\simeq m_a$. Once the axion field starts to move, the effective Lagrangian is given by,
\begin{equation}
 {\mathcal{L}}\supset \frac{1}{f_a}\partial_\mu \phi_a j^\mu;~~~ \j^\mu = 
 \sum_f \bar{\psi_f} \gamma^\mu \psi_f,
 \label{eq:coupling}
\end{equation}
where $\psi_{f}$ are SM lepton ($l_i$) and quark ($q_i$) doublets
and right-handed singlets of different flavors. Integrating the Eq.(\ref{eq:coupling})
by parts,
\begin{equation}
 {\mathcal{L}}\supset - \frac{1}{f_a} \phi_a \partial_\mu j^\mu
 \rightarrow \frac{\phi_a}{f_a} \frac{N_g}{8\pi^2} 
 \left( g_2^2 W_{\alpha \beta} \tilde{W}^{\alpha \beta} -
 - g_1^2 B_{\alpha \beta} \tilde{B}^{\alpha \beta}\right),
 \label{eq:coupling-ferm}
\end{equation}
where $N_g$ is the number of SM fermion generations. The 
electroweak field strength tensors are represented as $W_{\mu\nu}$
and $B_{\mu\nu}$.
The gauge 
couplings of the electroweak sector are represented as $g_1$
and $g_2$.  In Eq.(\ref{eq:coupling-ferm}) the anomaly equation
has been used. Considering the axion field is spatially homogeneous;
$\phi_a = \phi_a(t)$, the coupling presented in Eq.(\ref{eq:coupling-ferm}) transforms into an effective chemical potential
$\mu_{\rm eff}$ for the fermion number. The latter can be identified
by performing an integration by parts in Eq.(\ref{eq:coupling-ferm}),
\begin{equation}
 {\mathcal{L}}\supset\frac{\dot{\phi_a}}{f_a} j^0 = \mu_{\rm eff} n_F,
 \label{eq:chemical}
\end{equation}
where $\mu_{\rm eff } = \dot{\phi_a}/f_a$ and $j^0 = n_F = n_f- n_{\bar{f}}$. The non-zero effective chemical potential $\mu_{\rm eff}$ creates a shift in the energy levels of particles and antiparticles.
In thermal equilibrium, the minimum of the free energy is obtained for a
nonzero fermion-antifermion asymmetry, $n_F$,
\begin{equation}
 n_{f, \bar{f}}^{\rm eq} \sim T^3\left( 1 \pm \frac{\mu_{\rm eff }}{T},
 \right), ~~n_F^{\rm eq} =  n_{f}^{\rm eq} - n_{\bar{f}}^{\rm eq} \sim \mu_{\rm eff} T^2.
\end{equation}
The equilibrium number density of the equilibrium number density can be calculated taking the quantum-statistical effects for counting relativistic degrees of freedom in the classical Boltzmann approximations.
%
So the necessary condition for generating a lepton asymmetry is satisfied \cite{Kusenko:2014uta}. The mechanism does not depend on $CP$ violating phases or the heavy neutrino mass spectrum, unlike thermal leptogenesis.
Lepton number violation is mediated by the $\Delta L =2$ scattering
processes. The scalar triplets generate the $LL \leftrightarrow \bar{H}\bar{H}$ density rate $\gamma_{T_s}$ by $s$-channel exchange and $LH \leftrightarrow \bar{L}\bar{H}$ density rate 
 $\gamma_{T_t}$ by $t$-channel exchange. Here $H$ is a generic
 Higgs scalar representing $H_u/H_d$.
The process $LL \longleftrightarrow HH$ is in thermal equilibrium
\begin{equation}
 \Gamma_L|_{T= M_\Delta} = n^{\rm eq}_l \sigma_{\rm eff}  > H|_{T= M_\Delta},~ n^{\rm eq}_l = \frac{2}{\pi^2} T^3,
\end{equation}
where $ \sigma_{\rm eff}  = \langle \sigma_{\Delta L =2} v\rangle $ is the thermally averaged cross-section of 
${\Delta L =2}$ process. And,
\begin{equation}
  H|_{T= M_\Delta} = \sqrt{\frac{\pi^2 g_*}{90}} \frac{M^2_\Delta}{M_{\rm Pl}}
\end{equation}
Identifying 
\begin{equation}
\Gamma_L|_{T=M_\Delta} =  n^{\rm eq}_l \langle \sigma_{\Delta L =2} v \rangle \approx \frac{Y_\Delta^2 \mu^2}{M_\Delta},
\end{equation}
and using the expression for light neutrino mass in Eq.(\ref{eq:nu-mass-std}) $m_\nu \simeq Y_\Delta \frac{|\mu|v_{\rm ew}^2}{2 M^2_{\Delta}}$, 
the limit on the triplet Higgs mass results in 
\begin{equation}
 M_\Delta > 10^{12} {\rm GeV}.
\end{equation}

The other lepton number violating processes $LL\longleftrightarrow \Delta$
and $HH \longleftrightarrow \Delta$ must not co-exist, otherwise the 
lepton number asymmetry will be rapidly washed out. If both processes
are in equilibrium, the solutions corresponding to chemical potentials
are $\mu_L= \mu_{u_L} =\mu_H =0$. To maintain the lepton asymmetry in the plasma, the process $LL\longleftrightarrow \Delta$ must be efficient while
the process $HH \longleftrightarrow \Delta$ is out of equilibrium. The 
 latter condition can be represented as
\begin{equation}
 \Gamma_{ID}(HH \longleftrightarrow \Delta)|_{T= M_\Delta} < H|_{T= M_\Delta},
 \label{eq:gamma_ID}
\end{equation}
where 
\begin{equation}
  \Gamma_{ID}(HH \longleftrightarrow \Delta)|_{T= M_\Delta}\approx
 \Gamma_{D}(\Delta \rightarrow HH) \simeq \frac{\mu^2}{32\pi M_\Delta}.
\end{equation}
Using $v_\Delta \sim \frac{\mu v^2_{ew}}{2 M_\Delta^2}$, the condition in Eq.(\ref{eq:gamma_ID})
on $v_\Delta$ translates to
\begin{equation}
 v_\Delta \lesssim {\mathcal{O}}(1) {\rm GeV}.
\end{equation}

In a study \cite{Barrie:2021mwi,Barrie:2022cub}, the authors study 
the combination of the triplet Higgs and SM Higgs can
simultaneously give rise to a Starobinsky-like inflationary
epoch \cite{Starobinsky:1980te}. Efforts to use an axion with
a shift symmetry to ensure a prolonged slow-roll background evolution was studied in \cite{Pajer:2013fsa}.
Primordial inflation is produced by a
combination of $\sigma$ (nonminimally coupled to the scalar curvature) and the SM Higgs boson was explored in \cite{Ballesteros:2016euj}.
In the framework of {\it natural Inflation} \cite{Freese:1990rb}, Nambu-Goldstone
fields like axion can drive the inflation. But the results from Planck 2018 + BICEP/Keck(BK15) Polarization + BAO  data strongly disfavors the framework 
\cite{Stein:2021uge}. The alternate mechanism of axion as an inflaton candidate
has been studied in references \cite{Kim:2004rp} and \cite{DAmico:2017cda}. A model with chaotic inflation where the role of inflaton is played by the Higgs triplet in type II seesaw mechanism for generating the small masses of left-handed neutrinos was studied in Ref. \cite{Chen:2010uc}. In this work, we assume, during inflation, the axion remains spectator responsible for creating a chemical potential required for leptogenesis and the inflation is driven by another inflaton field. 

The requirement that the inflaton must not decay before the end of inflation implies:
\begin{equation}
 \Gamma_\phi \le H_{\rm inf},
\end{equation}
where $\Gamma_\phi$ is the decay width of the inflaton.
Planck measurements of the CMB anisotropies, combining information from the temperature and polarization maps and the lensing reconstruction put a constraint on the scale of 
inflation as reference $\frac{H_{\rm inf}}{M_{Pl}} < 2.5 \times 10^{-5}~ (95 \% ~{\rm CL})$ \cite{Planck:2018vyg}. We make use
of this bound in our numerical analysis for estimating the BAU 
from the leptogenesis via axion oscillation in the next subsection.
\subsection{Numerical analysis and results}
\label{subsec:num}
The successful implementation of triplet assisted leptogenesis requires that the triplet scalar mediated $\Delta L =2$ interactions must be in 
thermal equilibrium before the onset of axion oscillation
\begin{equation}
 \Gamma_L \gg H \gtrsim m_a,
\end{equation}
where 
\begin{equation}
\Gamma_L = n_l^{\rm eq} \sigma_{\rm eff},~ n_{l}^{\rm eq} = \frac{2}{\pi^2}T^3,
\label{eq:cond}
\end{equation}
where the equilibrium number density $n_{l}^{\rm eq}$
has been calculated by taking the quantum-statistical effects for counting relativistic degrees of freedom. 
Now the conditions given in Eq.(\ref{eq:cond}),
result in 
\begin{equation}
 T\sim T_L = \frac{\sqrt{g_*}
 }{\pi \sigma_{\rm eff} M_Pl} \sim 10^{13}~ 
 {\rm GeV};~~ m_a\sim \sigma_{\rm eff}~ T_L^3 \sim 10^8 ~{\rm GeV},
\end{equation}
where,
\begin{equation}
 \sigma_{\rm eff} \approx 
 \frac{3}{32 \pi} 
 \frac{\bar{m}^2}{v^4_{\rm ~ew}}
  \simeq 1\times 10^{-31}~ {\rm ~ GeV}^{-2};~~\bar{m}^2 = \sum^{3}_{i=1} m^2_i \approx \Delta m^2_{\rm atm} \simeq
  2.4 \times 10^{-3} {\rm eV^2}.
  \label{eq:sigma_eff}
\end{equation}
The dependence of $\sigma_{\rm eff}$ on light neutrino mass 
can be understood from the calculation of 
Feynman diagrams representing $\Delta L =1$ and $\Delta L =2$  processes using Feynman rules for the dimension-5 operator, $\kappa \propto m_\nu/v_{\rm ew}^2$ \cite{Strumia:2006qk, Nir:2007zq}. 
The solution to temperature is obtained by using energy density due to radiation,
\begin{equation}
 \rho_R = \frac{\pi^2 g_*}{3} T^4
\end{equation}
After the end of inflation, the temperature rises to the maximal value
\begin{equation}
 T_{\rm max} = 5\times 10^{13} {\rm ~GeV} \left( 
 \frac{\Gamma_\phi}{10^9 {\rm~ GeV}}\right)^{1/4}
 \left(\frac{H_{\rm inf}}{10^{11} {\rm~ GeV}}\right)^{1/2}.
\end{equation}
After that due to inflaton oscillation the the energy density scales
as matter and hence the temperature decreases. During reheating
the temperature drops as $T\propto R^{-3/8}$ until the radiation comes to dominate the energy density of the universe at time $t= t_{\rm rh} \simeq
\Gamma_\phi^{-1}$ when $\rho_R = \rho_\phi$ ($\rho_\phi$ is the energy density of inflaton) where the reheating temperature is ($T_{\rm rh} \simeq 0.3\sqrt{\Gamma_\phi~M_{\rm Pl}}$)
\begin{equation}
 T_{\rm rh} \simeq 2\times 10^{13} ~{\rm GeV} 
 \left( \frac{\Gamma_\phi}{10^9~ {\rm GeV}} \right)^{1/2}
\end{equation}
To get the final lepton asymmetry and hence the BAU the following
equations must be solved simultaneously,
\begin{eqnarray}
 \ddot{\phi_a} +3H \dot{\phi_a}&=& -\frac{\partial V(\phi_a)}
 {\partial \phi_a}
 \label{eq: BEs1}, \\
 \dot{\rho}_\phi +3 H\rho_\phi &= & -\Gamma_\phi \rho_\phi, \\
 \dot{\rho}_R +4 H\rho_R &= & \Gamma_\phi \rho_\phi, \\
 H^2 &=& \frac{\rho_{\rm tot}}{3 M_{Pl}^2}, \\
 \rho_{\rm tot} &\approx&  \rho_\phi+\rho_R, \\
 \dot{n}_L + 3H n_L &=& -\Gamma_L \left( n_L - n_L^{\rm eq}\right).
 \label{eq:BEs}
\end{eqnarray}
\begin{figure}
\centering
\begin{subfigure}{0.5\textwidth}
  \centering
  \includegraphics[width=1.1\linewidth]{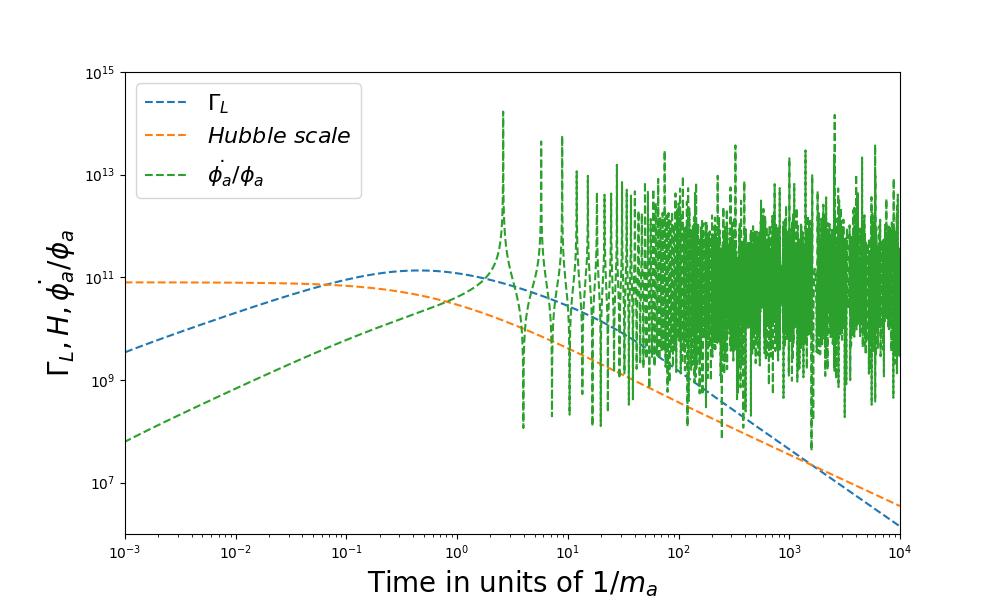}
  \caption{Evolution of $H, \Gamma_L$ and $\dot{\phi_a}/\phi_a$.}
  \label{fig:sub1}
\end{subfigure}%
\begin{subfigure}{0.5\textwidth}
  \centering
  \includegraphics[width=1.1\linewidth]{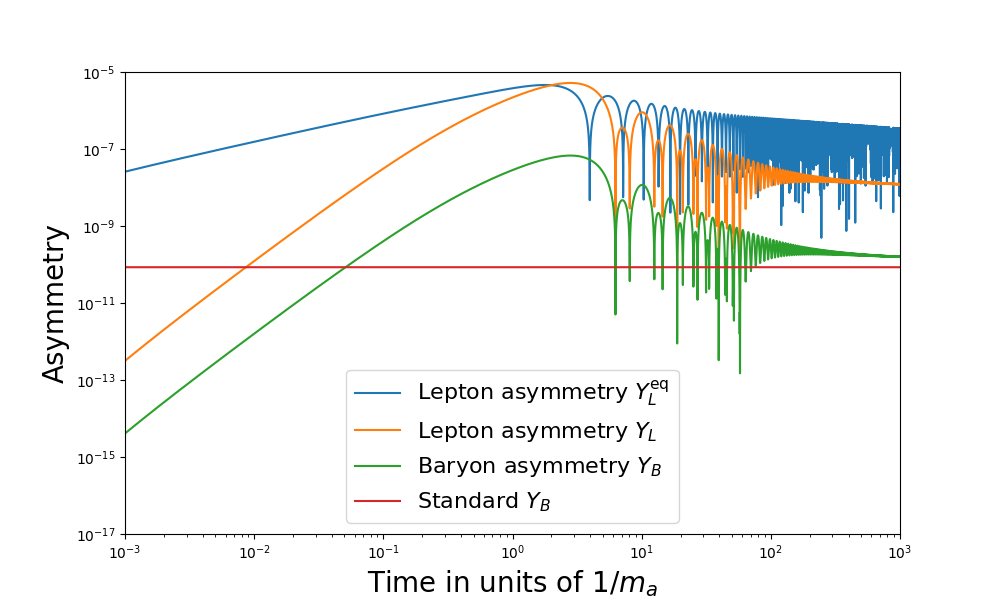}
  \caption{Evolution of lepton and baryon asymmetry.}
  \label{fig:sub2}
\end{subfigure}
\caption{The plots represent the evolution of various quantities of interest as 
a function of time in units of $1/m_a$. In the left panel of the figure 
the evolution of the Hubble parameter, $H$, the rate of 
$\Delta L =2$ process $\Gamma_L$ and the oscillation of the ALP as $\dot{a}/a$. In the right panel the evolution of equilibrium value of lepton asymmetry $Y_L^{\rm eq} = n_L^{\rm eq}/s$,  lepton asymmetry $Y_L =n_L/s$ and final baryon asymmetry $Y_B =n_B/s$. The red line is the reference line for the observed value of $Y_B = 8.75 \times
10^{-11}$. The numerical evolution of the quantities has been 
done for the benchmark values: $f_a = 3\times 10^{14}$ GeV, $m_a = 7\times 10^{10}$ GeV, $\Gamma_\phi = 5\times 10^{9}$ GeV and $H_{\rm inf} = 8\times 10^{10}$ GeV.}
\label{fig:test}
\end{figure}
Taking quantum-statistical effects for counting relativistic degrees of freedom, the equilibrium number density in the classical Boltzmann approximations
\begin{equation}
 n_L^{\rm eq}= \frac{4}{\pi^2} \mu_{\rm eff} T^2.
 \end{equation}
The lepton asymmetry is partly converted to baryon asymmetry by the
electroweak sphaleron processes. The final BAU of $\eta_B$ can be
calculated as,
\begin{equation}
 \eta_B= c_{\rm sph} \eta_L; ~~~\eta_L = \frac{n_L}{s},
\end{equation}
where $s$ is the entropy density of the Universe and $c_{\rm sph} = \frac{28}{79}$. The plots in Fig.(\ref{fig:test}) represent the evolution of various quantities of interest as 
a function of time in units of $1/m_a$. The numerical evolution of the quantities has been 
done by solving Boltzmann equations (\ref{eq: BEs1} - \ref{eq:BEs}) for the benchmark values: $f_a = 3\times 10^{14}$ GeV, $m_a = 7\times 10^{10}$ GeV, $\Gamma_\phi = 5\times 10^{9}$ GeV and $H_{\rm inf} = 8\times 10^{10}$ GeV. In the left panel of the figure 
the evolution of the Hubble parameter, $H$, the rate of 
$\Delta L =2$ process $\Gamma_L$ and the oscillation of the ALP as $\dot{\phi_a}/\phi_a$ are shown in orange, blue, and green colors respectively.
Similarly in the right panel the evolution of equilibrium value of lepton asymmetry $Y_L^{\rm eq} = n_L^{\rm eq}/s$,  lepton asymmetry $Y_L =n_L/s$ and final baryon asymmetry $Y_B =n_B/s$ are shown. The red line is the reference line for the observed value of $Y_B = 8.75 \times
10^{-11}$. The initial value of ALP and its  velocity are taken to be 
$\phi_a =f_a$ and $\dot{\phi_a} =0$. Since the equilibrium lepton number
density $Y^{\rm eq}_{L}$ depends on $\mu_{\rm eff} = \dot{\phi_a}/\phi_a$, it tracks the evolution of $\dot{\phi_a}$. Once $\Gamma_L$, 
the $\Delta L=2$
triplet scalar mediated scattering catches up being in equilibrium
the lepton asymmetry $Y_L$ and hence baryon asymmetry $Y_B$ start to
grow and as the $\Gamma_L$ process goes out of equilibrium the lepton
asymmetry freezes. Finally, due to the Sphaleron process, the 
lepton asymmetry gets converted to baryon asymmetry $Y_B \sim 8.7 \times
10^{11}$.

In our numerical analysis, we find the final lepton asymmetry is highly sensitive to $\Gamma_L$ through its
dependence on thermally averaged cross-section $\sigma_{\rm eff}/ \langle \sigma_{\Delta L =2} v \rangle$.
As per Eq.(\ref{eq:sigma_eff}), the latter is determined 
partly from the light neutrino mass $m_\nu$. In the model considered in the present work, the light neutrino mass 
matrix is generated through the type-II seesaw mechanism. The 
direct link between $m_\nu$ and triplet Yukawa coupling 
$Y_\Delta$ is now directly linked to the high-energy scale
phenomenon of leptogenesis. In the next section, we discuss 
the low-energy scale connection of $Y_\Delta$ in the lepton flavor violation processes.

\section{Lepton flavor violation}
\label{sec:lfv}
The generation of light neutrino masses and mixing generated through the seesaw mechanism can induce lepton flavor violation (LFV)
effects in charged lepton sector like $l_i\rightarrow l_j \gamma$.
The LFV effects are suppressed in models where SM is extended
with heavy particles responsible for the seesaw mechanism. 
The processes can be sizeable in the
minimal supersymmetric standard model (MSSM) embedded in GUTs
where they can be enhanced through the one-loop exchange of sleptons (and gauginos).  The LFV effects manifest from the 
Yukawa sector of the superpotential and the RGE effects 
that introduce off-diagonal elements of the slepton mass matrix.
The supersymmetric version of the model considered in this work
and the LFV effects can be compared with the model as studied in
Ref. \cite{Rossi:2002zb}.
%
%
%
The existence of LFV effects in the slepton sector can result in enhanced LFV processes in the charged lepton sector. For example the radiative decays 
$l_i\rightarrow l_j+ \gamma$ can be estimated as
\begin{equation}
{\rm BR} \left( l_i\rightarrow l_j+ \gamma \right) \simeq
\frac{\alpha^3}{G_F^2} \frac{|(m^2_{\bar{L}})_{ji}|^2}
{m_S^8} \tan^2\beta ~{\rm BR}\left( l_j \rightarrow l_i
\nu_j \bar{\nu_i}\right),
\end{equation}
where $\alpha$ and $G_F$ denote the fine structure and Fermi constant respectively, and $m_S$ is a typical SUSY mass particles inside the loops. It is useful to work with the quantities
\begin{eqnarray}
 R_{\tau\mu}\equiv \frac{{\rm BR} \left( \tau \rightarrow \mu  \gamma \right)}{ {\rm BR} \left( \mu \rightarrow e \gamma
 \right)} \simeq \left| \frac{\left( m^2_{\tilde{L}}\right)_{\tau\mu}}{\left( m^2_{\tilde{L}}\right)_{\mu e}}   
 \right|^2 \frac{{\rm BR} \left( \tau \rightarrow \mu \nu_\tau 
 \bar{\nu}_\mu \right)}{ {\rm BR} 
 \left( \mu \rightarrow e \nu_\mu \bar{\nu}_e
 \right)} \\
 R_{\tau e}\equiv \frac{{\rm BR} \left( \tau \rightarrow e \gamma \right)}{ {\rm BR} \left( \mu \rightarrow e \gamma
 \right)} \simeq \left| \frac{\left( m^2_{\tilde{L}}\right)_{\tau e}}{\left( m^2_{\tilde{L}}\right)_{\mu e}}   
 \right|^2 \frac{{\rm BR} \left( \tau \rightarrow e  \nu_\tau 
 \bar{\nu}_e \right)}{ {\rm BR} 
 \left( \mu \rightarrow e \nu_\mu \bar{\nu}_e
 \right)}
\end{eqnarray}
Between the GUT scale $M_{\rm GUT}$ and the seesaw scale $M_\Delta$ the RGE 
induced slepton mass matrix inherits the LFV effect as
\begin{equation}
 \left( m^2_{\tilde{L}}\right)_{\tau e} \simeq 
 - \frac{9 m_0^2 + 3a_0^2}{8\pi^2}\left( Y_\Delta^\dagger  Y_\Delta \right)_{ij} {\rm ln} \frac{M_{\rm GUT}}{M_\Delta},
\end{equation}
where $m_0$ and $A_0$ are the universal soft mass for scalars and universal trilinear coupling defined at the GUT scale.
The striking feature of the type-II seesaw mechanism allows to replace 
$\left( Y_\Delta^\dagger  Y_\Delta \right)_{ij}$ by $\left( m_\nu^\dagger m_\nu \right)_{ij}$ and hence the link between LFV effects and low-energy neutrino
parameters become direct. Using the the ratios $\frac{{\rm BR} \left( \tau \rightarrow \mu \nu_\tau 
 \bar{\nu}_\mu \right)}{ {\rm BR} 
 \left( \mu \rightarrow e \nu_\mu \bar{\nu}_e
 \right)} = 0.1737$ and $\frac{{\rm BR} \left( \tau \rightarrow e  \nu_\tau 
 \bar{\nu}_e \right)}{ {\rm BR} 
 \left( \mu \rightarrow e \nu_\mu \bar{\nu}_e
 \right)} = 0.1784$
the quantities of interest now can be written as
\begin{eqnarray}
R_{\tau \mu } &\simeq& 0.1737\times  \left| \frac{\left( m_\nu^\dagger m_\nu \right)_{\tau\mu}}{\left( m_\nu^\dagger m_\nu\right)_{\mu e}} \right|^2,\\
R_{\tau e } &\simeq& 0.1784 \times \left| \frac{\left( m_\nu^\dagger m_\nu \right)_{\tau e}}{\left( m_\nu^\dagger m_\nu \right)_{\mu e}}   
 \right|^2.
\end{eqnarray}
%
\section{Conclusion}
\label{sec:concl}
In this work, we have studied a model where the presence of a single 
triplet scalar accounts for light neutrino mass and baryogenesis via 
leptogenesis. In contrast to the standard thermal leptogenesis,
the initial lepton asymmetry is generated from $CPT$ violating
effect from the ALP which is sourced from the spontaneous 
breaking of $U(1)_{\rm PQ}$. The scale of PQ symmetry-breaking
is decoupled from the electroweak symmetry breaking at a scale
$\sim 10^{14}$ GeV. The PQ symmetry is broken before the end of inflation and the initial field value of ALP is uniform. The classical
evolution of the ALP induces a chemical potential for the fermion number due to the 
derivative coupling of the ALP with the electroweak field strength
tensors. The requirement that there be an external
source of baryon or lepton number violation process
in equilibrium is met by the presence
of a single triplet, scalar is suitable enough to assist the 
$\Delta L =2$ process and generate light neutrino mass via type-II
seesaw mechanism. This requires the reheat temperature  be $\simeq10^{13}$
GeV makes it a high-energy scale phenomenon. 
Unlike the case of thermal leptogenesis, where one needs at least two scalar triplets or one each of scalar triplet and right-handed singlet, a single triplet is enough to generate light neutrino mass and baryogenesis via leptogenesis. Moreover the $U(1)_{\rm PQ}$ extended SM provides a common link between neutrino mass generation and leptogenesis.  
We find the production of final lepton asymmetry is highly sensitive light neutrino mass $m_\nu$.  The 
proportionality relation between $m_\nu$ and triplet Yukawa coupling 
$Y_\Delta$ makes it now directly linked to the high-energy scale
phenomenon of leptogenesis. With this set up again the 
determining the flavor structure of the model from 
low-energy scale data is preserved. So the model provides 
an excellent link between low-energy and high-energy scale 
phenomena. Cosmology allows us to verify physical processes occurring at such high mass scales. 
For example, it was shown in Ref. \cite{Dev:2019njv} that if there is a coupling present 
between the SM Higgs and the ALP arising from a global $U(1)$ symmetry breaking, it induces a strong first-order phase transition, thereby producing stochastic gravitational waves that are potentially observable in current and future gravitational-wave detectors.
%
%
%
%
\appendix
\section{PQ charges}
\label{app:PQcharge}
The invariance of the Lagrangian and the scalar potential requires the 
constraints on the PQ charges as,
\begin{eqnarray}
 -X_l +X-e -X_d &=& 0 \\
 2X_l + X_\Delta &=& 0 \\
 2X_\Sigma -X_u - X_d &=& 0\\
 X_\sigma + X_u - X_\Delta -X_d &=& 0
\end{eqnarray}
The above equations can be solved in terms of $X_u$ and $X_d$ to get
\begin{eqnarray}
\nonumber
 X_l &=& - \frac{3X_u}{4} + \frac{X-d}{4}, \quad X_e =
 -\frac{3X_u}{4} + \frac{5 X_d}{4},\\ 
 X_\Delta &=& \frac{3X_u}{2} -\frac{X_d}{2}, \quad X_\sigma =
 \frac{X_u}{2} + \frac{X_d}{2}
 \label{eq:Xs}
\end{eqnarray}
The orthogonality of hypercharge and axion currents requires
\begin{equation}
 X_u^2 v_u^2 = X_d^2 v_d^2
 \label{eq:ortho}
\end{equation}
Using the conditions in equations (\ref{eq:Xs}) and (\ref{eq:ortho}),
all PQ charges can be determined up to an overall renormalization.
By choosing the renormalization condition
\begin{equation}
 X_\sigma =1,
\end{equation}
and defining $x \equiv v_u/v_d$ , the PQ charges are given by,
\begin{eqnarray}
 \nonumber
 X_u = \frac{2}{x^2 +1}, \quad X-d =\frac{2x^2}{x^2+1}, \quad X_l =
 \frac{x^2-3}{2(x^2+1)},\\
 X_e = \frac{5x^2 -3}{2(x^2 +1)}, \quad X_\Delta = \frac{3-x^2}{x^2+1}
\end{eqnarray}
\section{Scalar spectrum}
\label{sec:scalar}
With the electroweak VEVs switched off
\begin{equation}
 \frac{\partial V}{\partial f_a}
 = 2f_a \left( 2\lambda_3 f^2_a -\mu^2_3\right)
 \label{eq:min}
\end{equation}
Using Eq.(\ref{eq:min}) in expanding the scalar potential up to second order in the dynamical fields and with the unbroken SM symmetry, the scalar spectrum is given by: 
a real scalar SM singlet $\sigma^0$
\begin{equation}
 M^2_{\sigma^0} = 4\lambda_3 f_a^2.
\end{equation}
A real pseudoscalar SM singlet $\eta_\sigma^0$\begin{equation}
M^2_{\eta_\sigma^0} =0,
\end{equation}
corresponding to the zero-mass mode of the 
PQ-breaking field corresponding to the axion. A complex triplet with mass
\begin{equation}
 M^2_\Delta = \lambda_{\Delta 3}f_a^2 + 
 \mu^2_\Delta.
\end{equation}
The mass matrix of the complex doublets $H_u$ 
and $H_d$ in the basis $\left( H_u^*, H_u \right)$ 
\begin{equation}
 M^2_H =
 \begin{pmatrix}
  \lambda_{13} f_a^2 -\mu^2_1 & \lambda_5 f_a^2 \\
  \lambda^2_5 f^2_a & \lambda_{23} f_a^2 -\mu^2_2 
 \end{pmatrix}.
\end{equation}
Diagonalizing the above mass matrix, the corresponding eigenvalues are given as
\begin{equation}
 2 M^2_{u,d} = (\lambda_{13} +\lambda_{23})
 f^2_a -\mu^2_1 -\mu^2_2 \pm
 \sqrt{\left( (\lambda_{13} -\lambda_{23})f^2_a -\mu^2_1 +\mu^2_2\right)^2 + 4\lambda_5^2f^4_a}
  \end{equation}

\bibliographystyle{unsrt}

\end{document}